\documentclass[prl,twocolumn,amsmath,amsfonts,amssymb]{revtex4} 
\usepackage{epsfig,psfrag,graphicx,bm,color}
\pdfoutput=1

\begin{document}
\title{Dark matter interpretation of recent electron and positron data}

\author{Lars Bergstr\"om}
\email{lbe@physto.se}
\author{Joakim Edsj\"o}
\email{edsjo@physto.se}
\author{Gabrijela Zaharijas}
\email{gzah@physto.se}
\affiliation{The Oskar Klein Centre for Cosmoparticle Physics, Department of Physics, Stockholm University, AlbaNova, SE - 106 91 Stockholm, Sweden}



\begin{abstract}
We analyze the recently released Fermi-LAT data on the sum of
electrons and positrons. Compared to a conventional, pre-Fermi, background model, a surprising excess in the several hundred GeV
range is found and here we analyze it in terms of dark matter models. We also compare with
newly published results from PAMELA and HESS, and find models giving very good fits to these data sets as well. If this dark matter interpretation is correct, we also predict the possibility of a sharp break in the diffuse gamma ray spectrum coming from final state radiation.
\end{abstract}

\maketitle


\section{Introduction}

Recently, the question of dark matter in the universe has been put into focus, as data from a new generation of cosmic ray detectors have been tentatively interpreted in terms of indirect signatures of dark matter annihilation (or decay).
In particular, although the data on antiprotons from the PAMELA satellite \cite{pamela_pbar}, and the mid-latitude data on diffuse gamma rays from Fermi \cite{non-gev-excess} show no hints of an ``exotic'', extra component, the positron fraction measured by PAMELA \cite{pamela_positrons} and the sum of electrons and positrons by ATIC \cite{atic} definitely show an excess compared to standard 
background predictions. At any rate, these indications of an extra component has sparked a remarkable number of theoretical papers, both concerning dark matter explanations (for recent reviews, see \cite{hooper,lberev}) and other astrophysical sources such as pulsars (e.g., \cite{pulsars}).

With the very recent release of Fermi-LAT \cite{fermi}  data on the sum of electrons and 
positrons \cite{fermicre}, the experimental precision has been increased in an impressive way. Although these high-quality data do not confirm the peak claimed by ATIC, they indeed also indicate an excess compared to the expected background as predicted in the so-called conventional, pre-Fermi GALPROP model \cite{galprop}.  In this paper we show, however, that the dark matter annihilation
interpretation seems to fit even better with the new data. We compute confidence regions of the parameters needed in the dark matter sector to fit the data, and we show examples of the remarkably high quality fits obtained by some of the best-fit models. Interestingly, the models that fit Fermi also reproduce the break indicated at high energy by the HESS collaboration \cite{hess}. One should keep in mind though, that the pulsar interpretation seems still to be a possibility worthy of further studies
\cite{dario}.

We point out two features of viable models, which had already been 
inferred in some other work by 
trying to fit previous data and checking other bounds (in particular, from gamma-ray \cite{gammacon,cirelli,cuoco} and radio emission \cite{synchro,bbbet,cholis}). 
Of course, the previous work tried to fit the ``bump'' claimed by the ATIC experiment, which is not seen in the new Fermi data. This means that smaller boost factors are actually required, and also the best fit mass may be increased. 
First, our solutions for the dark matter mass are in the 1--4 TeV range, which is indeed where the needed cross section enhancement 
by the Sommerfeld factor is operative \cite{hisano}.  
Another mechanism giving a boost is the possible
non-uniformity of the dark matter distribution within the nearest kpc of the solar system, or a combination of both, see, e.g., \cite{brun,kuhlen}, where possible boost
factors much higher than needed in this work are obtained.

Second, we find that the measured distribution seems to favour direct dark matter annihilation to $\mu^+\mu^-$, with very little of $e^+e^-$ or $\tau^+\tau^-$, thus 
pointing to an unusual underlying particle physics model. However, if the annihilation is first into low-mass states which subsequently decay into leptons, both the absence of $\tau^+\tau^-$ (for kinematical reasons) 
and the occurrence of $\mu^+\mu^-$ 
can be naturally explained, e.g.\ if the intermediate particle is scalar or 
pseudoscalar, by helicity arguments. 

\section{Dark matter and propagation models}

We will consider two different types of dark matter models: i) leptonic models with direct annihilation to leptons, and as we will see, preferably to
 $\mu^+\mu^-$, and ii) models annihilating to a light scalar (or pseudo-scalar or vector) decaying predominantly to leptons, where we focus on the Arkani-Hamed et al.\ (AH) type of
 models \cite{arkani-hamed} and the Nomura and Thaler (N) type of models \cite{nomura}. We use {\sf DarkSUSY} \cite{darksusy} to calculate the positron fluxes for these models and for the case of the AH type models, we have modified {\sf DarkSUSY} to allow for the scalar (vector) decaying directly to $e^+e^-$. The propagation model we use is the one from \cite{eb} where we use the propagation parameters of 
the MED model of \cite{delahaye}. One should note though, that at the highest energies, which is of most interest in fitting the Fermi and
 HESS \cite{hess} data, only the energy loss parameter $\tau_0$ plays a significant role and we choose
$\tau_0=10^{16}$ s/GeV$^2$ as a default value. For the dark matter halo model, to be conservative, we choose an isothermal sphere with a local dark matter density of $\rho_0=0.3$ GeV/cm$^3$. However, our results are not
 very sensitive to the actual halo model, apart from the local density $\rho_0$. For all classes of models, we have set the annihilation cross section to $(\sigma v) = 3\times 10^{-26}$ cm$^{3}$ s$^{-1}$. Note that all the fluxes scale with $\rho_0^2$ and at high energies (all energies of interest here), they also scale with $\tau_0$. With a boost factor from substructure and/or Sommerfeld enhancements, $B_F$, we can then define an overall enhancement factor,
\begin{equation}
E_F = \left( \frac{\rho_0}{0.3~{\rm GeV\,cm^{-3}}} \right)^2 \left( \frac{\tau_0}{10^{16}~\rm s\, GeV^{-2}} \right) B_F
\label{eq:ef}
\end{equation}
which to a very good approximation will scale our fluxes up (or down). The enhancement factors 
we find below thus includes the Sommerfeld enhancement arising in the AH and N type of models. For both these latter models, we use the benchmark models AH1--AH4 and N1--N5 as defined in \cite{bbbet}.

We will also discuss other constraints, like constraints from gamma rays \cite{gammacon,cirelli,cuoco} and synchrotron radiation \cite{synchro,bbbet,cholis}.

\section{Data and backgrounds}

In our analysis, we will include the recent data from Fermi \cite{fermicre} and HESS \cite{hess} for the summed spectrum of  electrons and positrons, and 
Pamela data on the positron fraction \cite{pamela_pbar}. 
We will not include ATIC \cite{atic} in our fits, as those data do not seem compatible with Fermi (although an excess is indeed also found by Fermi, the peak structure seen by ATIC is not). For HESS, we consider energies up to 2 TeV. Since the background model we use agrees with the HESS data (below 2 TeV) within $2 \sigma$, we use HESS data only to place an upper limit on our models, by requiring that our signal + background does not overproduce electrons at the HESS energies.

For the background estimate, we use the conventional 
GALPROP \cite{galprop} model (i.e.\ not specifically adjusted to fit the recent data; see also further discussion in the next Section), as given in \cite{fermicre}.

As both data and background estimates have systematic
 uncertainties, we have some freedom of adjusting their relative normalizations. It is a non-trivial task to do this in a fully consistent way, but looking at the background estimate and the data it seems that the HESS data and the background estimate is slightly higher than the Fermi data. To take this normalization 
uncertainty into account, we have arbitrarily rescaled the HESS data and the background estimate by 0.85, which is within the expected systematic uncertainties of Fermi, HESS and the background estimates.

\section{Results of fits}

\begin{figure*}[t!]
\includegraphics[width=0.32\textwidth]{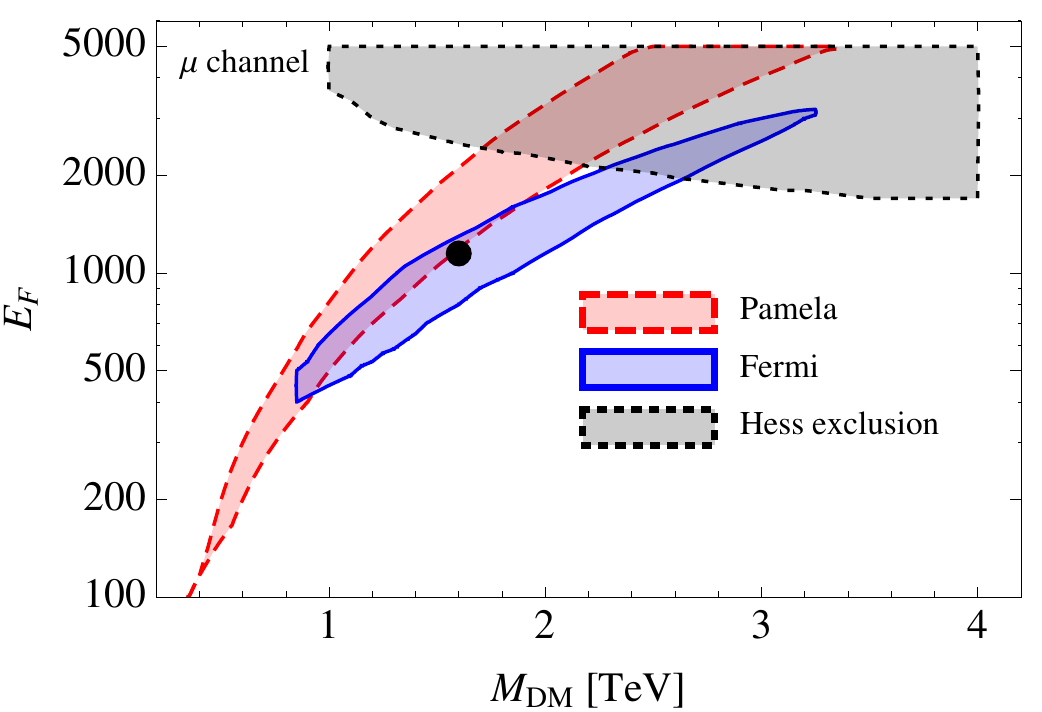}
\includegraphics[width=0.32\textwidth]{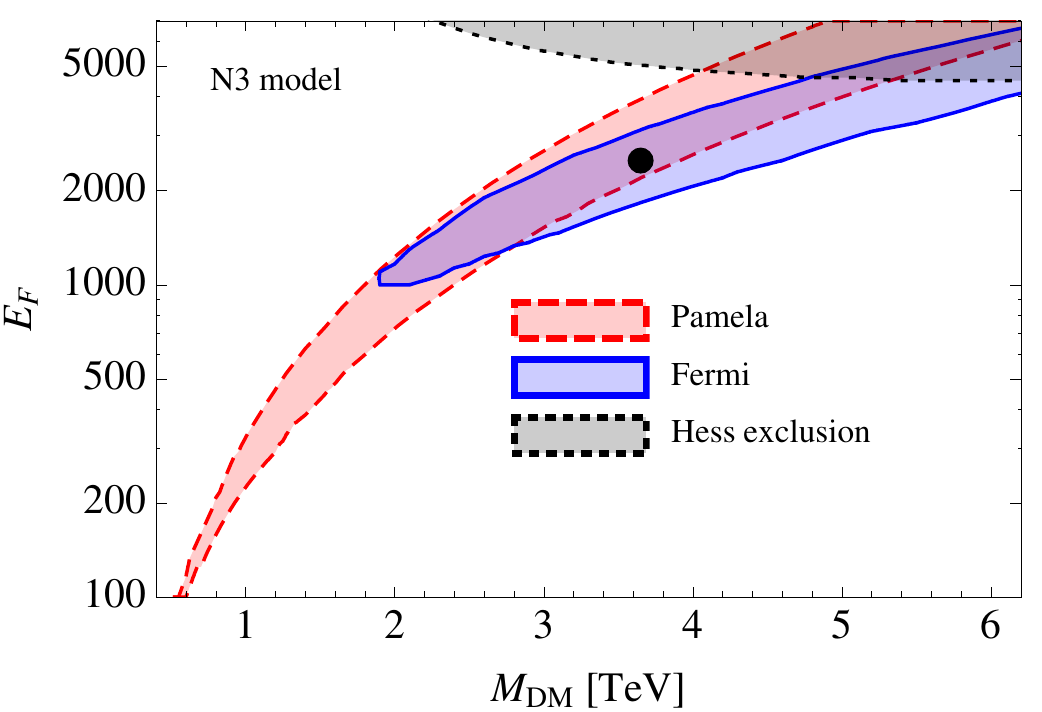}
\includegraphics[width=0.32\textwidth]{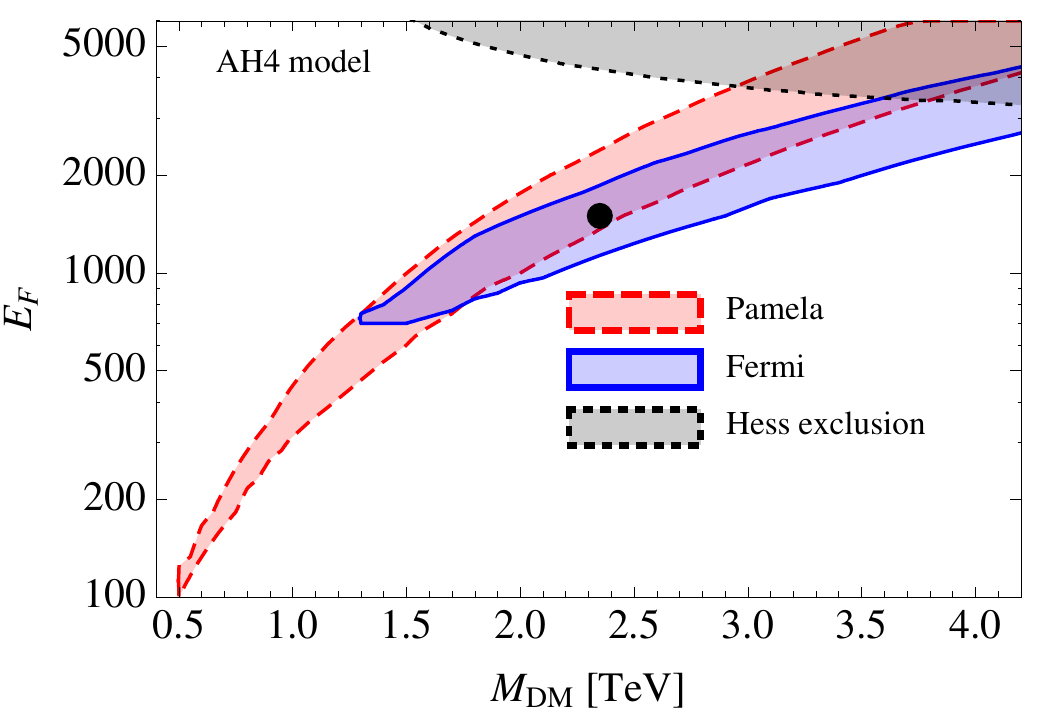}
\caption{The $2\sigma$ contours in the enhancement factor - mass plane for a) annihilation to $\mu^+\mu^-$, b) the Nomura-Thaler model N3 and c) the Arkani-Hamed et al.\ model AH4. The contours are shown for PAMELA and Fermi, whereas the HESS data is only used as an upper limit. The black dot is the example model shown in Fig.\protect\ref{fig:spec}.}
\label{fig:boost}
\end{figure*}

\begin{figure*}[t!]
\includegraphics[width=0.32\textwidth]{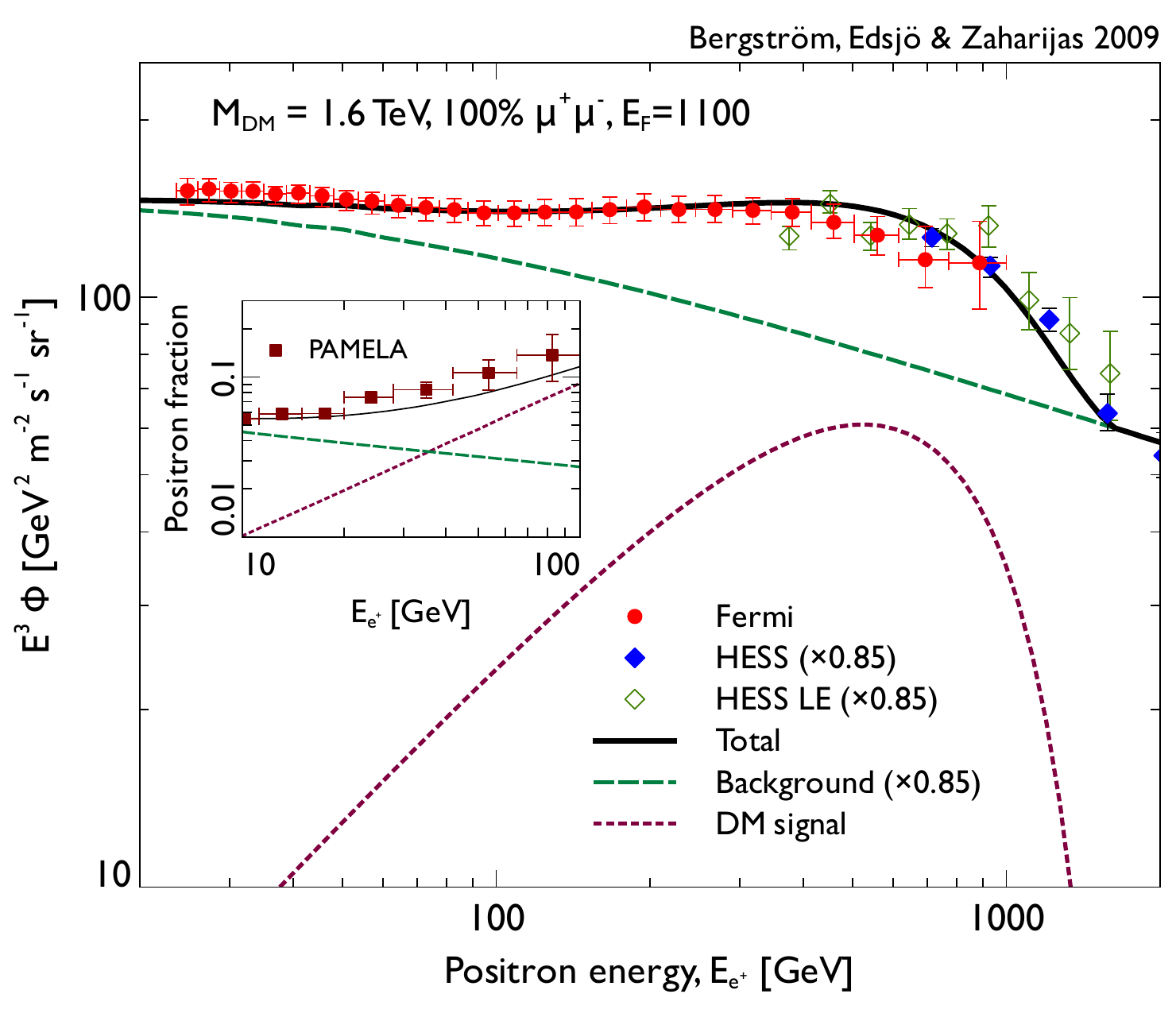}
\includegraphics[width=0.32\textwidth]{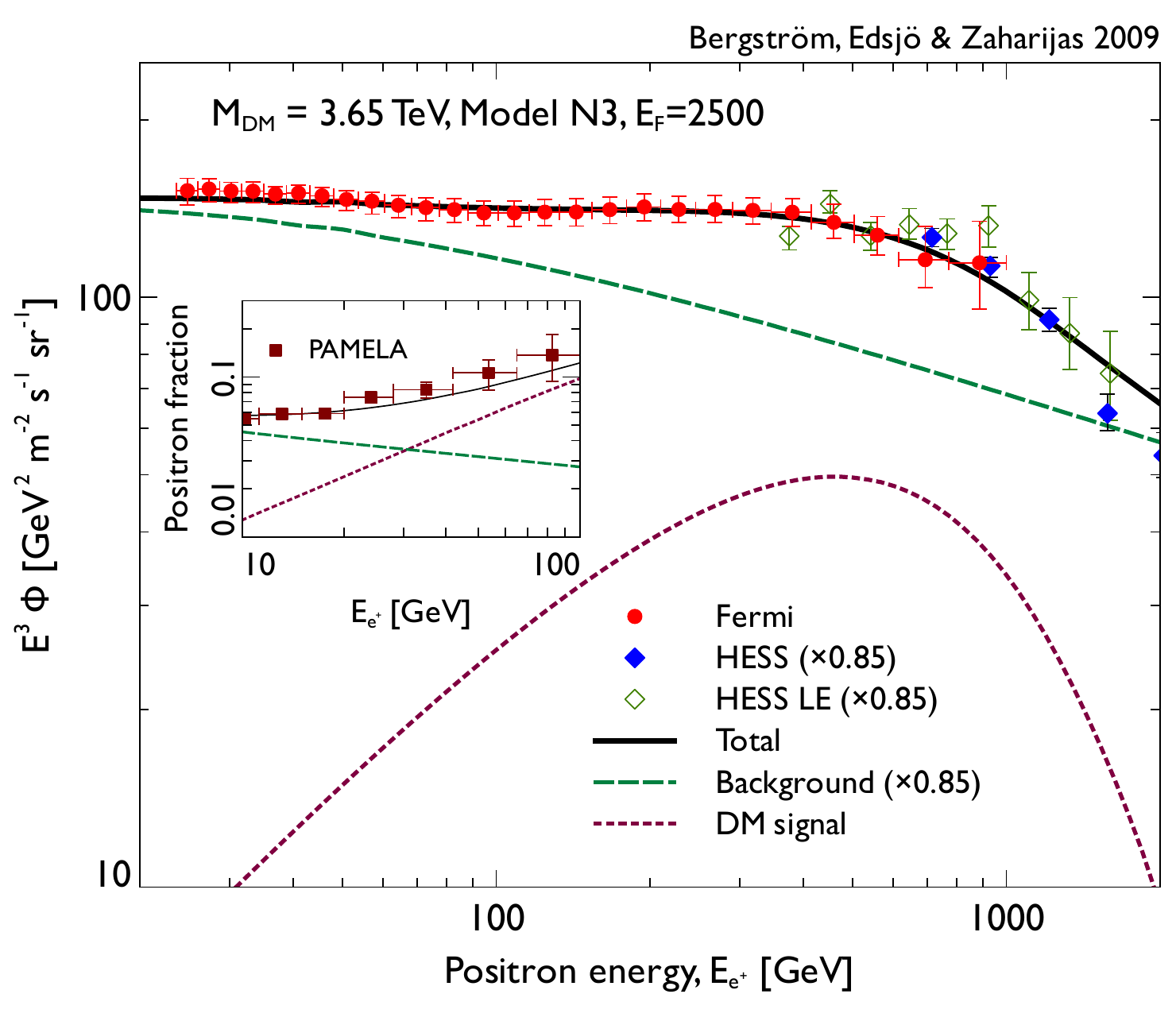}
\includegraphics[width=0.32\textwidth]{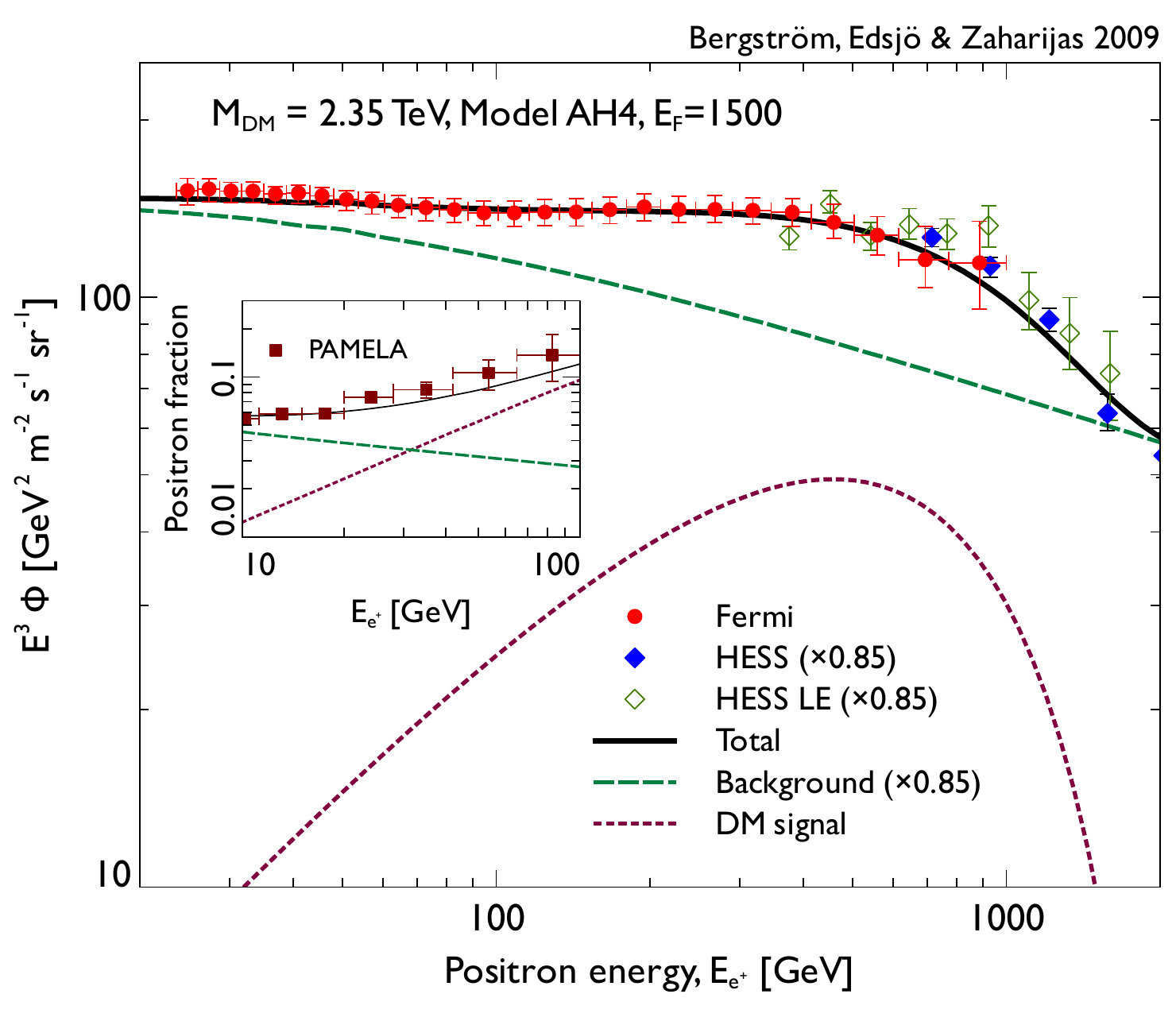}
\caption{Spectra for examples of good fit models in \protect\ref{fig:boost}. The signal and background are shown for electrons ($e^+ + e^-$) together with Fermi \cite{fermicre} and HESS data \cite{hess,hessle}. The HESS data and the background model has been rescaled with a factor 0.85. In the inset, the positron fraction as measured with PAMELA is shown together with the predicted signal for the same model.}
\label{fig:spec}
\end{figure*}

We are now ready to perform our dark matter fits to the data. First we consider the leptonic models. It turns out that annihilation directly to $e^+e^-$ typically gives a too pronounced peak at the WIMP mass $M_{DM}$. Tau leptons typically give a too soft spectrum and too many gamma rays from $\pi^0$ decays from hadronic tau decays \cite{cuoco,cholis}. Hence, we will for these leptonic models focus on models annihilating only to $\mu^+\mu^-$. It can be argued that this is unnatural (and not expected) for most dark matter candidates, but will come back to this question when we discuss the Nomura-Thaler and Arkani-Hamed type of models. 

In Fig.~\ref{fig:boost}a we show the required enhancement factor versus mass assuming 100\% annihilation to $\mu^+\mu^-$. In \ref{fig:boost}a, the contours are shown for fits to Fermi and PAMELA data within $2 \sigma$ and the exclusion region which is $2 \sigma$ above the HESS data. As can be clearly seen, we typically get rather large regions with good fits to the data. The boost factors are of the order of 1000 and typically, slightly larger boost factors would be needed to get good fits also to the PAMELA data on the positron fraction. The quality of the fit is shown in Fig.~\ref{fig:spec}a, where we show the spectrum for an example of a good fit model (the dot in Fig.~\ref{fig:boost}a)). As can be seen, the excess in the Fermi data is very well-fit with this dark matter model and we get a reasonable (but a bit low) fit to the PAMELA data as well.\footnote{In Fig.~\ref{fig:spec} we also show the recently announced HESS low-energy data \cite{hessle}, even if these data are not included in our fits.}

For the leptonic model, it is rather unnatural to assume that annihilation only goes to $\mu^+\mu^-$. However, for the AH \cite{arkani-hamed} and N \cite{nomura} type of models, this follows naturally from the mass of the scalar being above $2m_\mu$, but below $2m_\tau$. In the N models, the light scalar (actually a pseudo-scalar) decays only to $\mu^+\mu^-$, whereas in the AH models, the decay can go either purely to $\mu^+\mu^-$ or to a mixture of $e^+e^-$ and $\mu^+\mu^-$, depending on the nature of the scalar. Not surprisingly, the N models and the AH4 model (which has a scalar that decays 100\% to $\mu^+\mu^-$) give the best fit to the electron and positron data and we therefore focus on these. In Figs.~\ref{fig:boost}b and \ref{fig:spec}b, we show the $2\sigma$ contours and an example spectrum for a good fit N3 type of model where we have only kept the enhancement factor $E_F$ and WIMP mass $M_{DM}$ as free parameters (all the other N1--N4 models considered in \cite{bbbet} give the same positron fluxes and are thus equivalent in this respect).

In Figs.~\ref{fig:boost}c and \ref{fig:spec}c, we show the same $2\sigma$ contours and example spectrum for the AH4 model. Even if the fits to the data for model N3 and AH4 are not as excellent as for the muon channel model, they are still good, and of course it is more natural to get predominantly muons in these models. Another advantage of these types of models is that it is natural to get the required boost factors which enter the total enhancement factor (Eq.~(\ref{eq:ef})) from the Sommerfeld enhancement.

Of course, one has to investigate other constraints on these kinds of models, especially given the rather large boost factors needed to fit the electron and positron data. For the leptonic model, we can use the constraints in \cite{synchro}, where the gamma ray and radio constraints, primarily towards the galactic centre and dwarf spheroidals were investigated. For Einasto or NFW profiles, the best fit models are excluded due to gamma rays from the galactic centre. However, for less steep profiles, like an isothermal sphere, our best fit models are not excluded by these data. 

For the N and AH models, constraints from gamma rays and radio (including final state radiation photons) were investigated in \cite{bbbet}. The same conclusion holds for these models, if the halo profile is an Einasto or NFW profile (or steeper), the models are already excluded. However, for shallower halo profiles, like an isothermal sphere, the models are still viable. One should note that the electron and positron fluxes discussed in this paper are not very dependent on the choice of halo profile, so the best-fit models derived here, would be more or less the same for an NFW profile instead of the isothermal profile we used in our analysis.

Given the large amounts of high-energy electrons and positrons injected into the galaxy with these models, it is also fair to wonder about secondary radiation from inverse Compton scattering on the interstellar radiation field \cite{cholis,zhang,cuoco,cirelli}. In \cite{cuoco} it is  concluded that models annihilating to $\mu^+\mu^-$ are at tension with EGRET data and that Fermi will be able to probe these models. Given the new Fermi data, lower boost factors are needed than those assumed in \cite{cuoco}, so the tension with EGRET data is less severe. However, Fermi should still be able to probe these models. For the N3 and AH4 model, we get very similar constraints \cite{cholis} and these are also viable with a shallow halo profile.

One should also note that
we have chosen to work with a rather standard halo and diffusion model, but it is rather straightforward to rescale our results via the enhancement factor introduced in Eq.~(\ref{eq:ef}). Note that the dependence on $\rho_0$ and $\tau_0$ in Eq.~(\ref{eq:ef}) is a very good approximation for high energies. For lower energies (i.e.\ the PAMELA range), it is more involved as the positrons at these energies have propagated rather far. Keeping the signal fixed at higher energies, it is possible to move the signal from dark matter up at lower energies by having a larger significant diffusion region (by having a larger diffusion zone half height and a larger diffusion coefficient). Increasing $\tau_0$ will also increase the fluxes at low energies slightly more than the linear relation in Eq.~(\ref{eq:ef}) as positrons then sample a larger (and partly denser) region in the galaxy. These effects are more pronounced for steeper halo profiles, like a Navarro-Frenk-White \cite{nfw} profile, than the isothermal profile considered here. In either case, these effects are rather small, and Eq.~(\ref{eq:ef}) remains a very good approximation also for the PAMELA energies, even for steeper halo profiles.

\section{Conclusions and discussion}
\label{sec:disc}

We see that it is certainly possible to fit the excess seen in the Fermi data with a dark matter model. We have here used a `standard' diffusion model and a `standard' dark matter halo, but one should keep in mind that we need to invoke boost factors (either from substructure and/or a Sommerfeld enhanced annihilation cross section) of the order of $10^3$ for this to work. For the dark matter models annihilating directly to leptons, we also prefer models annihilating predominantly to $\mu^+\mu^-$ as the $e^+e^-$ channel gives too many hard electrons and the $\tau^+\tau^-$ channel gives too soft electrons (and possibly too many gamma rays). Hence, for these models, we would need to invoke a rather unnatural selection of the $\mu^+\mu^-$ channel. However, for the Sommerfeld enhanced models with light scalars, we naturally get annihilation to only (or at least predominantly) $\mu^+\mu^-$ through annihilation to light scalars which then (for kinematical reasons) decay mostly to muons. These models typically require a slightly larger mass and enhancement factor, but on the other hand in these models the enhancement factor from Sommerfeld enhancement is expected, a we get the annihilation to muons `for free'.

One should also note that we have not varied the halo or propagation model parameters, which could help in getting an even better fit also to PAMELA data, as discussed above.

We have here used the conventional, pre-Fermi, conventional GALPROP \cite{galprop} model for our background estimate. This model should most certainly be revised given the new data, but we leave this for future work.

A very nice property of DM models, as was recently pointed out \cite{kuhlen} and as verified in this work, is the relative model independence of the shape of the DM spectrum at low energies. The case for a dark matter interpretation of the new surprising data is in our view getting stronger, although we have to await even more data, plausibly using gamma-rays, to settle the issue.  

As a final note, for the very best fit model that annihilates directly into muons (or, but to a lower extent,
models with spin-0 bosons decaying to muons), there may in fact exist a 
unmistakeable signature that would make it uniquely different from the pulsar interpretation (or other backgrounds), 
namely the direct emission of final state radiation (FSR) in the annihilation process. Although the 
authors of \cite{cuoco} unfortunately did not add that contribution to the total flux in their Fig.~1, 
they show its magnitude as a dashed red line. It is obvious that a very striking feature should appear 
-- a slow rise of an excess in the $E^2dN_\gamma/dE$ curve up to 100-300 GeV where the rise 
should suddenly be much steeper as the FSR $dN_\gamma/dE\propto 1/E$ sets in, i.e.\ a linear rise 
when multiplied by $E^2$ (see Fig.~1 in \cite{bbbet}). However, the existence and exact location of this feature will depend on 
the rather poorly known starlight energy density near the galactic centre, and also the detailed spatial 
distribution of the DM halo. It will be very interesting to see if Fermi-LAT can find 
(or rule out) such a behavior when they analyze the diffuse gamma radiation from the region nearer the galactic
centre in more detail. At mid-latitudes and at lower energies such an excess is neither present
\cite{non-gev-excess} nor predicted to exist in the models discussed
here.

\section*{Acknowledgements}

We acknowledge the hard work by our experimental colleagues in the 
Fermi LAT Collaboration, which has provided us with data of 
unprecedented quality and precision.  We thank members of the 
collaboration for important comments and feedback, especially Jan
Conrad, Stefano Profumo, Luca Baldini and Nicola Omodei.

The support of the Swedish Research Council (VR) is greatly acknowledged.
We wish to thank Emiliano Sefusatti for help with producing some of the figures.



\begin{thebibliography}{99}
\bibitem{pamela_pbar}
  O.~Adriani {\it et al.},
  Phys.\ Rev.\ Lett.\  {\bf 102}, 051101 (2009)
  [arXiv:0810.4994 [astro-ph]].
\bibitem{non-gev-excess} G. Johannesson, for the Fermi Collaboration, talk presented at the AAS Meeting, Long Beach, CA, Jan. 2009, to be published in the Proceedings.
\bibitem{pamela_positrons} 
  O.~Adriani {\it et al.}  [PAMELA Collaboration],
  arXiv:0810.4995 [astro-ph].

\bibitem{atic}
  J.~Chang {\it et al.},
  Nature {\bf 456} (2008) 362.
\bibitem{hooper}
  D.~Hooper,
  arXiv:0901.4090 [hep-ph].

\bibitem{lberev}
 L.~Bergstr\"om,
  arXiv:0903.4849 [hep-ph].

\bibitem{pulsars}
A.K. Harding and R. Ramaty, Proc. 20th ICRC, Moscow {\bf 2}, 92-95 (1987);
F.A. Aharonian, A.M. Atoyan and H.J. V\"olk, Astron. Astrophys. 
{\bf 294} L41 (1995);
I. B\"{u}sching, O.C. de Jager, M.S. Potgieter and C. Venter,
Astrophys. J. {\bf 78}, L39-L42 (2008);
  S.~Profumo,
  arXiv:0812.4457 [astro-ph];
  D.~Malyshev, I.~Cholis and J.~Gelfand,
  arXiv:0903.1310 [astro-ph.HE].

\bibitem{fermi} The Fermi-LAT collaboration, see e.g. 
  W.~B.~Atwood {\it et al.}  [LAT Collaboration],
  arXiv:0902.1089 [astro-ph.IM].
\bibitem{fermicre}
A.A.~Abdo {\it et al.}, Phys.\ Rev.\ Lett.\ {\bf 112} (2009) 181101. [arXiv:0905.0025].

\bibitem{galprop}
The GALPROP computer code, see home page http://galprop.stanford.edu and, e.g., 
  A.~W.~Strong, I.~V.~Moskalenko and O.~Reimer,
  Astrophys.\ J.\  {\bf 613}, 962 (2004)
  [arXiv:astro-ph/0406254].
\bibitem{hess}
  F.~Aharonian {\it et al.}  [H.E.S.S. Collaboration],
  Phys.\ Rev.\ Lett.\  {\bf 101}, 261104 (2008)
  [arXiv:0811.3894 [astro-ph]].


\bibitem{dario} D.~Grasso {\it et al.}, to appear.

\bibitem{gammacon}
  M.~Cirelli, M.~Kadastik, M.~Raidal and A.~Strumia,
  Nucl.\ Phys.\  B {\bf 813}, 1 (2009)
  [arXiv:0809.2409 [hep-ph]].

\bibitem{cuoco}
  E.~Borriello, A.~Cuoco and G.~Miele,
  arXiv:0903.1852 [astro-ph.GA].

\bibitem{cirelli}
  M.~Cirelli and P.~Panci,
  arXiv:0904.3830 [astro-ph.CO].


\bibitem{synchro}
  G.~Bertone, M.~Cirelli, A.~Strumia and M.~Taoso,
  arXiv:0811.3744 [astro-ph].
  
\bibitem{cholis}
  I.~Cholis, G.~Dobler, D.~P.~Finkbeiner, L.~Goodenough and N.~Weiner,
  arXiv:0811.3641 [astro-ph].
  
\bibitem{bbbet}
  L.~Bergstr\"om, G.~Bertone, T.~Bringmann, J.~Edsj\"o and M.~Taoso,
  arXiv:0812.3895 [astro-ph]; Phys. Rev. D, in press.

\bibitem{hisano}
  J.~Hisano, S.~Matsumoto and M.~M.~Nojiri,
  Phys.\ Rev.\ Lett.\  {\bf 92}, 031303 (2004)
  [arXiv:hep-ph/0307216].

\bibitem{brun}
  P.~Brun, T.~Delahaye, J.~Diemand, S.~Profumo and P.~Salati,
  arXiv:0904.0812 [astro-ph.HE];
  J.~Bovy,
  arXiv:0903.0413 [astro-ph.HE].

\bibitem{kuhlen}
  M.~Kuhlen and D.~Malyshev,
  arXiv:0904.3378 [hep-ph].

\bibitem{arkani-hamed}
N.~Arkani-Hamed, D.P.~Finkbeiner, T.R.~Slatyer and N.~Weiner,
[arXiV:0810.0713].

\bibitem{nomura} Y.~Nomura and J.~Thaler,
[arXiv:0810.5397].

\bibitem{darksusy}
  P.~Gondolo, J.~Edsj\"o, P.~Ullio, L.~Bergstr\"om, M.~Schelke and E.~A.~Baltz,
  JCAP {\bf 0407}, 008 (2004)
  [arXiv:astro-ph/0406204].


\bibitem{eb}
  E.~A.~Baltz and J.~Edsj\"o,
  Phys.\ Rev.\  D {\bf 59}, 023511 (1999)
  [arXiv:astro-ph/9808243].

\bibitem{delahaye}
  T.~Delahaye, F.~Donato, N.~Fornengo, J.~Lavalle, R.~Lineros, P.~Salati and R.~Taillet,
  arXiv:0809.5268 [astro-ph].

\bibitem{hessle}
F.~Aharonian {\it et al.}, [H.E.S.S. Collaboration], arXiv: 0905.0105.

\bibitem{zhang}
 J.~Zhang, X.~J.~Bi, J.~Liu, S.~M.~Liu, P.~F.~Yin, Q.~Yuan and S.~H.~Zhu,
  arXiv:0812.0522 [astro-ph].

\bibitem{nfw}
J.F.~Navarro, C.S.~Frenk and S.D.M.~White, Astroph.\ J.\ {\bf 490}, 493 (1997), [arXiv: astro-ph/9611107].

\end{thebibliography}
\end{document}